\documentclass[12pt]{article}
\pdfoutput=1
\usepackage{amssymb,graphicx}
\usepackage{epstopdf}
\usepackage{amsmath,amsfonts}
\usepackage{epsfig,xcolor,url}

\definecolor{DarkGreen}{rgb}{0,0.5,0}

\begin{document}

\begin{flushright}
INR-TH/2016-047
\end{flushright}
{\let\newpage\relax

%%%%%%%%%%%%%%%%%%%%%%%%%%%%%%%%%%%%%%%
%%%%%%%%%%%%%%%%%%%%%%%%%%%%%%%%%%%%%%%

\sloppy

\title{\bf General quadrupolar statistical anisotropy: Planck limits}
\author{S.~Ramazanov$^{a}$\footnote{{\bf e-mail}: sabir.ramazanov@gssi.infn.it}\;, 
G.~Rubtsov$^{b}$\footnote{{\bf e-mail}: grisha@ms2.inr.ac.ru}\;,
M.~Thorsrud$^{c}$\footnote{{\bf e-mail}: mikjel.thorsrud@hiof.no}\;, 
F.~R.~Urban$^{d}$\footnote{{\bf e-mail}: federico.urban@kbfi.ee}
\\
$^a$ \small{\em Gran Sasso Science Institute (INFN), Viale Francesco Crispi 7, I-67100 L'Aquila, Italy}\\ 
$^b$ \small{\em Institute for Nuclear Research of the Russian Academy of Sciences,}\\
\small{\em Prospect of the 60th Anniversary of October 7a, 117312 Moscow, Russia}\\
$^c$ \small{\em Faculty of Engineering, \O stfold University College,}\\
\small{\em P.O. Box 700, 1757 Halden, Norway}\\ 
$^d$\small{\em National Institute of Chemical Physics and Biophysics, R\"avala 10, 10143 Tallinn, Estonia}}
\maketitle

\begin{abstract}

Several early Universe scenarios predict a direction-dependent
spectrum of primordial curvature perturbations.  This translates into
the violation of the statistical isotropy of cosmic microwave
background radiation.  Previous searches for statistical anisotropy mainly
focussed on a quadrupolar direction-dependence characterised by a
\emph{single} multipole vector and an overall amplitude $g_*$.
Generically, however, the quadrupole has a more complicated geometry
described by \emph{two} multipole vectors and $g_*$.  This is the
subject of the present work.  In particular, we limit the amplitude
$g_*$ for different shapes of the quadrupole by making use of Planck
2015 maps.  We also constrain certain inflationary scenarios which
predict this kind of more general quadrupolar statistical
anisotropy. 

\end{abstract}}

%%%%%%%%%%%%%%%%%%%%%%%%%%%%%%%%%%%%%%%%%%%%%%%%%%%%%%%%%%%%%%%%%%%%%%%%%%%%%%
\section{Introduction and main results}\label{sec:int}
%%%%%%%%%%%%%%%%%%%%%%%%%%%%%%%%%%%%%%%%%%%%%%%%%%%%%%%%%%%%%%%%%%%%%%%%%%%%%%

With the release of the Cosmic Microwave Background (CMB) maps obtained with the Planck satellite, a string of properties of primordial scalar perturbations 
has been established with unprecedented accuracy. In particular, possible deviations 
from Gaussianity and adiabaticity are now subject to quite severe constraints, 
while exact scale-invariance of the primordial power spectrum is excluded at more than $5\sigma$ C.L.~\cite{Ade:2015xua, Ade:2015lrj}. 
These observations show no departure from the simplest idea of single scalar slow roll inflation, 
while narrowing the window for many alternative scenarios.  

Along with Gaussianity and adiabaticity, large-scale statistical isotropy (SI) of the 
Universe is a basic prediction of standard inflationary cosmology. This stems 
from the spin-0 nature of the inflaton and the isotropy of the (quasi)de Sitter 
space-time~\cite{Wald:1983ky} resulting in the rotational invariance of the inflaton field's correlation functions; together with the isotropy 
of the background metric during radiation- and matter-dominated stages, 
this implies the SI of CMB temperature fluctuations $\delta T ({\bf n})$. 
In particular, this means that the variance $\langle \delta T^2 ({\bf n}) \rangle$ is independent 
of the direction ${\bf n}$ in the sky. This can be paraphrased in harmonic 
space as the diagonality of the covariance matrix. 

Although SI is a robust prediction of inflation, there are examples of models which 
break this symmetry. At the level of primordial scalar perturbations $\zeta$, this 
amounts to saying that the power spectrum is direction-dependent, 
\begin{equation}
\label{powergen}
{\cal P}_{\zeta} ({\bf k}) ={\cal P}_{\zeta} (k) \left( 1+ {\cal Q} ({\bf k}) \right) \; .
\end{equation}
It is convenient to expand the function ${\cal Q} ({\bf k})$ in a series of spherical harmonics~\cite{Pullen:2007tu}, 
\begin{equation}
\label{Qspher}
{\cal Q} ({\bf k}) =\sum_{LM} q_{LM} (k) Y_{LM} (\hat{{\bf k}}) \; .
\end{equation}
where $L$ is an even number\footnote{The absence of multipoles with odd $L$ in the expansion of the power spectrum follows from the symmetry ${\cal P} ({\bf k}) ={\cal P} (-{\bf k})$. 
In turn, the latter is guaranteed by the commutativity of the curvature perturbation field $\langle [\zeta ({\bf x}), \zeta ({\bf y})] \rangle=0$.}. Generically, the coefficients $q_{LM}$ may have a dependence on the cosmological wavenumber $k$. 

Note that Eq.~\eqref{powergen} does not cover all the possible cases of SI violation. Indeed, there have been numerous 
anomalies seen in WMAP and Planck data~\cite{Bennett:2010jb, Ade:2015hxq, Schwarz:2015cma}: low amplitude of the quadrupole, quadrupole-octupole 
alignment, lack of large angular correlations~\cite{Bennett:2003bz, deOliveira-Costa:2003utu, Bielewicz:2005zu, Copi:2005ff}; hemispherical asymmetry/dipole modulation of the CMB sky~\cite{Eriksen:2007pc, Hansen:2008ym}; the cold spot~\cite{Vielva:2003et}, etc. 
These (at least some of them) may
hint towards the existence of a preferred direction in the sky and hence a breaking of SI. Although these anomalies were the primary trigger for considering 
direction-dependent primordial spectra, there seems to be no direct link between the former and the latter\footnote{For instance, the quadrupole-octupole alignment or the dipole modulation of the CMB temperature imply the existence of non-zero correlations between CMB temperature coefficients with multipole numbers $l$ and $l'=l \pm 1$. At the same time, the direction-dependence as in Eq.~\eqref{powergen} leads to correlations between multipoles differing by an even number.}. 
Therefore, we do not discuss the CMB anomalies in what follows. In fact, the spectra~\eqref{powergen} have been 
motivated in a string of early Universe scenarios, some of which we list below.

Depending on the type of directional-dependence, i.e., the function 
${\cal Q} ({\bf k})$, one can classify the anisotropies of the early Universe as follows: 
\begin{itemize}

\item Axisymmetric quadrupole. 
In this case there exists a reference frame where all but one of the $q_{LM}$ coefficients can be turned to zero, leaving only $q_{20}$. This type of statistical anisotropy (SA) is the 
most widely discussed in the literature. Besides its simplicity, 
it is a common prediction of anisotropic models of inflation as well as of some alternatives. 
The former include the historically first Ackermann--Carroll--Wise model~\cite{Ackerman:2007nb}, 
scenarios with vector curvaton~\cite{Dimopoulos:2008yv, dim2, dim3}, scenarios with a gauge field coupled to 
waterfall fields in hybrid inflation~\cite{Yokoyama:2008xw} and the generic class of models with a single Maxwellian gauge field 
coupled to the inflaton~\cite{Watanabe:2009ct, Watanabe2, Soda:2012zm, Bartolo:2012sd, Lyth:2013sha}. Alternatives to inflation are represented  
by models of the (pseudo)Conformal Universe~\cite{Hinterbichler:2011qk, Creminelli:2012qr,  Libanov:2015iwa}, i.e., conformal rolling scenario~\cite{Rubakov:2009np, Libanov:2010nk, Libanov:2011bk} and Galilean genesis~\cite{Creminelli:2010ba}. The common denominator of those scenarios is 
the existence of a \emph{single} long-ranged vector, which is responsible for SI breaking\footnote{More generically, abandoning the rotational invariance of the background in the early Universe leads to non-zero primordial SA, see, e.g.,~\cite{Gumrukcuoglu:2007bx, Ashoorioon}.}. Rotations with respect to this vector leave the primordial power spectrum intact. 
Thence, the axial symmetry of the quadrupole.

\item General quadrupole. This type of prediction is less widely discussed in literature. It follows, for example, from inflationary 
scenarios with {\it multiple} Maxwellian fields coupled to the inflaton~\cite{Yamamoto:2012tq, Thorsrud:2013kya, Thorsrud:2013mma} 
as well as from the (pseudo)Conformal Universe~\cite{Creminelli:2012qr, Libanov:2010nk}. 
In this case a single parameter is not enough to describe the anisotropy, 
as the axial symmetry is broken; a second quantity---the quadrupole shape $\chi$---needs to be taken into account ~\cite{Thorsrud:2013kya, Thorsrud:2013mma}. The parameter 
$\chi$ measures the deviation from axial symmetry. The particular choice $\chi=0$ corresponds to leaving the latter intact. We will sometimes refer to the shape $\chi$ as the 'angle' for reasons which will become clear in Section~2. SA of general quadrupolar form is the main focus of this work.

\item Higher multipoles. This prediction about SA arises in some versions of the (pseudo)Conformal Universe~\cite{Libanov:2011hh}; we will not deal with higher multipoles here.

We see how SA---at least in principle---could be a useful 
tool for discriminating among inflationary models as well as alternative frameworks. 

\end{itemize}

%Tests of primordial statistical isotropy were carried out relatively recently. Motivated by the 
%ACW model, they have been applied to constrain axially symmetric directional dependence. 
%The result obtained by making use of WMAP5 data was quite unexpected: a strong signal of 
%statistical anisotropy was found in the $W$ frequency band, which stood out 
%at about $9\sigma$ CL~\cite{Groeneboom:2008fz}. However, the direction of the quadrupole was eventually found to be aligned with the poles of the ecliptic plane~\cite{Groeneboom:2009cb, Hanson:2009gu}. Together with other features---frequency dependence of 
%the signal and non-confirmation in other experiments---the signal 
%must be rather attributed to a systematical error rather than non-trivial cosmology. 
%Indeed, in Ref.~\cite{Hanson:2010gu}, it was shown that the signal disappears upon inclusion 
%of the beam asymmetries. Furthermore, the effects of non-circular beams 
%have been implemented in WMAP9 and Planck maps. 

So far, most of the data analysis focussed on axisymmetric quadrupolar SA.  In that case, the power spectrum takes the 
form ${\cal P}_{\zeta} ({\bf k}) \propto \left( 1+ g_* \cos^2 \theta \right) $ in the appropriate reference frame. Here $g_*$ is the amplitude of 
SA and $\theta$ is the angle between the wavevector ${\bf k}$ and the preferred direction in the sky. The bounds obtained by exploiting the quadratic maximum likelihood estimators~\cite{Hanson:2009gu} are given by~~\cite{Ade:2015lrj} (see also Ref.~\cite{Ade:2015hxq}),
\begin{equation}
\label{planck2015}
-0.010<g_*<0.019 \qquad 68\%~\mbox{C.L.}
\end{equation}
for Planck~2015 data and 
$-0.046<g_*<0.048$ ($68\%~\mbox{C.L.}$) for WMAP9 data~\cite{Ramazanov:2013wea}.\footnote{Earlier releases of the WMAP data exhibited a strong axisymmetric quadrupolar SA with a direction aligned with the poles of ecliptic plane~\cite{Hanson:2009gu, Groeneboom:2008fz, Groeneboom:2009cb, Ramazanov:2012za}. This SA, however, was an artifact of using circular beam transfer functions~\cite{Hanson:2010gu}. Consequently, the signal of SA disappeared in the WMAP9 and Planck data upon including 
the non-circular beam effects~\cite{Ramazanov:2013wea,kim}. See Ref.~\cite{Das:2014awa} for more details. } Planck collaboration extended the analysis so that to include the possible 
$k$-dependence of the amplitude $g_*$~\cite{Ade:2015lrj}. In many cases, these constraints imply very stringent 
limits on the intrinsic parameters of the anisotropic early Universe scenarios.

In the present paper we search for the signatures of the general quadrupolar SA in the cosmological data for $k$-independent $q_{2M}$, see Eq.~\eqref{Qspher}, using Planck~2015 maps. 
Following Ref.~\cite{Rubtsov:2014yua}, we consider the data provided at $143$\,GHz and $217$\,GHz, and their cross-correlation. When formulating the final constraints, we stick to the cross-correlated data as the cleanest probe of SA.

The non-observation of any departures from SI bounds the amplitude of the
general quadrupole $g_*$ defined analogously to the case of the axisymmetric quadrupole. 
See Section~\ref{sec:est} for an exact definition. Notably, the data demonstrate different level of agreement with
different quadrupole shapes. This is clear from the resulting constraints on $g_*$, summarised in Table~\ref{Table1}, as a function of the shape quantified by the parameter $\chi$. We see a slight, but not statistically significant, preference towards negative 
amplitudes $g_*$. That tendency, prominent for sufficiently small
angles $\chi$, vanishes at larger values of $\chi$, see Section~3 for discussions. 

% This conclusion is somewhat at odds with the result~\eqref{planck2015} obtained by the Planck collaboration 
%for the case of the axisymmetric quadrupole ($\chi=0$). 
%The discrepancy, however, is not surprising and can be attributed to
%the different subsets of the Planck~2015 data used. See Section~3 for discussion. 

 For comparison purposes we also test the general quadrupole with Planck~2013 data. The latter, however, turn out to be insensitive to the shape of the quadrupole. As a result, 
the final constraints on the amplitude $g_*$ are independent on the angle $\chi$. These constraints on $g_*$ match very well the limits of Ref.~\cite{kim} (deduced specifically for the axisymmetric 
quadrupole).

Generically, in early Universe models (be it inflation or its competitors), the amplitude $g_*$ and the angle $\chi$ are not genuine model parameters, but they are random variables with distributions determined by the parameters specific for each scenario. 
This is a common situation in anisotropic inflationary scenarios with vector fields and in the (pseudo)Conformal Universe. 
Thus, a separate analysis is required in order to limit those models. We perform this analysis in Section~4 with a focus on inflationary scenarios which comprise (a collection of) Maxwellian fields non-minimally coupled to the inflaton~\cite{Watanabe:2009ct, Bartolo:2012sd, Thorsrud:2013kya, Thorsrud:2013mma}. In that case, SA is sourced by the infrared modes of these Maxwellian (gauge) vector fields, which follow a Gaussian distribution 
with a dispersion that grows linearly with the duration of inflation (in terms of the number of e-folds)~\cite{Bartolo:2012sd, Lyth:2013sha}. Therefore, the number of e-folds can be constrained from the 
non-observation of SA. In Section~4, we also briefly revisit the 
limits on some versions of the (pseudo)Conformal Universe.

This paper is organised as follows.  In Section~\ref{sec:est}, we
discuss the parametrisation of SA.  In Section~3, we assess the
sensitivity of the Planck data to the new parameter $\chi$. We
constrain inflationary scenarios with multiple vector fields from the
non-observation of SA in Section~4. Our constraining scheme there is
independent on the results of Sections~2 and~3. Therefore, the reader interested
only in those models may go directly to Section~4. 

%%%%%%%%%%%%%%%%%%%%%%%%%%%%%%%%%%%%%%%%%%%%%%%%%%%%%%%%%%%%%%%%%%%%%%%%%%%%%%
\section{Parametrisation}\label{sec:est}
%%%%%%%%%%%%%%%%%%%%%%%%%%%%%%%%%%%%%%%%%%%%%%%%%%%%%%%%%%%%%%%%%%%%%%%%%%%%%%

%%%%%%%%%%%%%%%%%%%%%%%%%%%%%%%%%%%%%%%
%%%%%%%%%%%%%%%%%%%%%%%%%%%%%%%%%%%%%%%
 
Upon choosing an appropriate coordinate system, 
one can write a general quadrupole as~\cite{Thorsrud:2013mma}
\begin{equation}
\label{genparametr}
\sum_M q_{2M} Y_{2M} (\hat{{\bf k}})=  \sqrt{\frac{16\pi}{45}}g_* \left\{ Y_{20} (\tilde{\vartheta}, \tilde{\varphi}) \cos \chi -
\frac{1}{\sqrt2} \left[ Y_{2,1} (\tilde{\vartheta}, \tilde{\varphi})-Y_{2,-1} (\tilde{\vartheta}, \tilde{\varphi})\right] \sin \chi \right\} \; .
\end{equation}
Here $g_*$ is the quadrupolar amplitude, and $\chi$ is an extra parameter (angle) which measures the departure from an axisymmetric quadrupole. 
Notice that in the case $\chi=0^{\circ}$ we come back to the usual axial symmetry. The angles $\tilde \vartheta$ and $\tilde{\varphi}$ correspond to the 
direction of the cosmological mode ${\bf k}$ in the new coordinate system. 
The orthonormal basis of this coordinate system is given by the triad of vectors 
$({\bf a}, {\bf b}, {\bf c})$ aligned with the axis $O\tilde{x}$, $O\tilde{y}$ and $O\tilde{z}$, respectively. 

The representation~\eqref{genparametr} of a general quadrupole has a particularly clear meaning in terms of multipole vectors~\cite{Copi:2003kt}. Up to a constant 
factor, we write
\begin{equation}
\nonumber 
\sum_M q_{2M} Y_{2M} (\hat{{\bf k}}) \propto ({\bf v}^{(2,1)} \cdot \hat{{\bf k}})  ({\bf v}^{(2,2)} \cdot \hat{{\bf k}}) - \frac{1}{3} {\bf v}^{(2,1)}\cdot{\bf v}^{(2,2)}   \; ,
\end{equation}
where ${\bf v}^{(2,1)}$ and ${\bf v}^{(2,2)}$ are mutipole vectors. As we show in the Appendix, the multipole vector representation reduces 
to the form~\eqref{genparametr} provided that one of the vectors is aligned with the 
$z$ axis of the new reference frame, while the other is lying in the $O\tilde{x}\tilde{z}$ plane. In the special case when two multipole vectors coincide, 
one recovers the axisymmetric quadrupole. The freedom of choosing the reference frame\footnote{Namely, two coordinate systems are associated with 
two multipole vectors. Two more coordinate systems are obtained by simultaneously changing the signs of the multipole vectors, see the
Appendix for details.} does not introduce any 
ambiguity in the quantities $g_*$ and $\chi$: they are uniquely defined in the region 
$-\infty<g_*< +\infty$ and $0^{\circ} \leq \chi \leq
  90^{\circ}$\footnote{In the special case of $\chi=90^{\circ}$, only $|g_*|$ is uniquely defined. This has no physical consequences since the difference between quadrupoles with  positive and negative $g_*$ smoothly goes to $0$ as $\chi\rightarrow90^\circ$, see Section~\ref{ssec:results}.}.

For practical purposes it is convenient to express the coefficients $q_{2M}$ in terms of the amplitude $g_*$ and the shape $\chi$, 
\begin{equation}
\label{q2mg} 
q_{2M} =\frac{8\pi g_*}{15} Y^{*}_{2M} ({\bf c}) \cos \chi +c^{*}_{2M} \sin \chi \; ,
\end{equation}
where 
\begin{equation}
\label{c2M}
c_{2M}=\frac{(8\pi)^{3/2} g_*}{9} \sum_{M'} Y^{*}_{1M'} ({\bf c}) Y^{*}_{1,-M'-M} ({\bf a})  
\left( \begin{array}{ccc} 
1 & 1 & 2\\
M' & -M-M' & M
\end{array} \right) \; .
\end{equation}
With the Planck data at hand, one may reconstruct the coefficients
$q_{2M}$ and confront observations with the coefficients calculated
for the given parameters  $g_*$ and $\chi$ using Eqs.~\eqref{q2mg}
and~\eqref{c2M}.

Notice that although we are always considering two parameters (viz, $g_*$ and $\chi$), 
an observer looking for imprints in the CMB will have to deal with three additional numbers determining the actual orientation of the quadrupole in the sky (namely, three angles fixing the orthornomal vectors ${\bf a}$ and ${\bf c}$). 
However, the theory typically tells nothing regarding the directions
of these vectors, which should thus be drawn from uniform
distributions. Marginalising over these three extra parameters one is
left with the amplitude $g_*$ and $\chi$ alone.
\section{Data analysis}\label{ssec:qlm}
%%%%%%%%%%%%%%%%%%%%%%%%%%%%%%%%%%%%%%%

\subsection{Tools and methods}

Conventionally, one employs quadratic maximum likelihood (QML) estimators to derive the $q_{LM}$ coefficients from CMB maps. Originally designed in Ref.~\cite{Hanson:2009gu}, they yielded the most stringent limits on SA to date~\cite{Ade:2015lrj, Ade:2015hxq, kim}. We slightly modify the QML estimators to make them appropriate for the cross-correlation analysis~\cite{Rubtsov:2014yua}. That is, we consider estimators of the form
\begin{equation}
\label{qlm}
q^{ij}_{LM}=\sum_{L'M'} ({\bf F}^{ij})^{-1}_{LM;L'M'} (h^{ij}_{L'M'} -\langle h^{ij}_{L'M'} \rangle ) \; ,
\end{equation}
where 
\begin{equation}
\label{hlm}
h^{ij}_{LM}=\sum_{ll';mm'}\frac{1}{2}i^{l'-l}C_{ll'} B^{LM}_{lm;l'm'} \bar{a}^{i}_{l,-m}  \bar{a}^{j}_{l'm'} \; .
\end{equation}
Here $\bar{a}^{i}_{lm}$ are related to standard CMB temperature coefficients $\hat{a}^{i}_{lm}$ by 
\begin{equation}
\label{inv}
\bar{a}_{lm} =\left({\bf C}^{iso}\right)^{-1}_{lm;l'm'} \hat{a}_{l'm'} \; .
\end{equation}
The upperscripts $i$ and $j$ denote a particular frequency band ($143$\,GHz or $217$\,GHz in our case); ${\bf C}^{iso}$ is a covariance (including the noise), which corresponds to {\it statistically isotropic} cosmological signal. The coefficients 
$C_{ll'}$ in Eq.~\eqref{hlm} are given by
\begin{equation}
\label{cll}
C_{ll'}=4\pi  \int d \ln k \Delta_l (k) \Delta_{l'} (k)  {\cal P}_{\zeta} (k) \; ,
\end{equation}
where ${\cal P}_{\zeta} (k)$ is an isotropic primordial power spectrum and $\Delta_l (k)$ is a transfer function. The structure constants $B^{LM}_{lm;l'm'}$ are expressed 
via the Wigner 3j-symbols
\begin{equation}
\nonumber 
B^{LM}_{lm;l'm'}=(-1)^M \sqrt{\frac{(2L+1)(2l+1)(2l'+1)}{4\pi}}
\left (
\begin{array}{ccc} 
L & l & l'\\
0 & 0 & 0
\end{array} 
\right ) \left (
\begin{array}{ccc} 
L & l & l'\\
M & m & -m'
\end{array} 
\right ) \; .
\end{equation}
The estimators~\eqref{qlm} are normalised by the Fisher matrix ${\bf F}^{ij}$ given by
\begin{equation}
\nonumber 
F^{ij}_{LM;L'M'} \equiv \langle h^{ij}_{LM} (h^{ij}_{L'M'})^{*} \rangle -\langle h^{ij}_{L'M'} \rangle 
\langle (h^{ij}_{L'M'})^{*} \rangle  \; .
\end{equation}
%The QML estimators are advantageous for our purposes because they are unbiased and have minimal variance. 
%As encoded in their name, they maximise the log-likelihood of the 
%observed CMB for the case $i=j$ in the approximation of small statistical anisotropy, $|q_{LM}| \ll 1$. 
In the homogeneous noise and full sky approximations, the Fisher matrix can be written in analytic form~\cite{Hanson:2009gu, Rubtsov:2014yua} as
\begin{equation}
\label{fish}
F^{ij}_{LM;L'M'} = \sum_{ll'} \frac{(2l+1)(2l'+1)}{16\pi}\frac{C^2_{ll'} \left( C^{tot,i}_l C^{tot,j}_{l'} +\tilde{C}^{ij}_l \tilde{C}^{ij}_{l'} \right)}{\left(C^{tot,i}_l \right)^2 \left( C^{tot,j}_{l'} \right)^2} \left (
\begin{array}{ccc} 
L & l & l'\\
0 & 0 & 0
\end{array} 
\right )^2 \delta_{LL'} \delta_{MM'} \; .
\end{equation}
Here $C^{tot,i}_l=C^{i}_l +N^{i}_l$, where 
$C^{i}_l$ and $N^i_l$ are the primordial and noise angular spectra derived from the $i$th band, respectively; $\tilde{C}^{ij}_l=C^{tot, i}_l$ for $i=j$ and 
$\tilde{C}^{ij}_l=C_l$ otherwise. 
%\begin{equation}
%F_{LM;L'M'} \approx  F^{unmasked}_L\int d{\bf n} W({\bf n}) Y^{*}_{L,-M} ({\bf n}) Y_{L',-M'} ({\bf n}) \; , 
%\end{equation}
Given an incomplete sky coverage, we proceed with a slight modification of the Fisher matrix, 
\begin{equation}
\nonumber 
F^{ij}_{LM;L'M'}  \rightarrow f_{sky} \cdot F^{ij}_{LM; L'M'} \; ,  
\end{equation}
where $f_{sky}$ is the unmasked fraction of the sky.

\vspace{0.2cm}

We followed the steps described below to derive $q_{LM}$ coefficients from the data and from Monte Carlo (MC) simulated maps:

\begin{itemize}
\item We obtain the temperature coefficients $\hat{a}_{lm}$ from Planck
  2015 data corresponding to 29 months of High Frequency Instrument
  (HFI) observations at frequencies $143$\,GHz and
  $217$\,GHz~\cite{HFI_maps}. We employ the HFI \url{GAL40} Galactic
  plane mask (\url{HFI_Mask_GalPlane-apo0_2048_R2.00.fits}) and HFI
  point sources mask (\url{HFI_Mask_PointSrc_2048_R2.00.fits}). The
  unmasked fraction of the sky is $f_{sky}=40.1\%$.

\item From $\hat{a}_{lm}$, we calculate the $\bar{a}_{lm}$ coefficients defined by Eq.~\eqref{inv}. Following Ref.~\cite{Hanson:2009gu}, we employ 
multigrid preconditioners~\cite{smith} at this step to reduce computational cost. 

\item The $C_{ll'}$ coefficients given by Eq.~\eqref{cll} are evaluated
  with the \emph{CAMB} package~\cite{Lewis:1999bs}. 

\item We compute the Wigner 3j-symbols and provide summations in Eqs.~\eqref{hlm} and~\eqref{fish} using {\it gsl}~\cite{gsl} and {\it slatec}~\cite{slatec} libraries. 
The summations run over the range of multipole numbers $l=[2, 1600]$. For
$l>l_{max}=1600$, the observed signal is dominated by the instrumental noise. Note that before the noise begins to dominate the Fisher matrix scales as $F \sim l^2_{max}$ with the maximal multipole number $l_{max}$ used in the analysis. Therefore, taking $l_{max} \ll 1600$ would lead to a 
large loss of sensitivity for SA. 

\item Finally, averaging in
  Eq.~\eqref{qlm} is performed by repeating the procedure for 100
  Monte-Carlo generated statistically isotropic maps. The latter are constructed using Full Focal Plane
  simulations for CMB and noise maps~\cite{Planck_MC} coadded with the
  \emph{SMICA} foreground map (\url{HFI_CompMap_Foregrounds-smica_2048_R2.00.fits})~\cite{Comp_separation}.
\end{itemize}

\subsection{Results \label{ssec:results}}

 \begin{figure}[tb!]
\begin{center}
\includegraphics[width=0.22\columnwidth,angle=-90]{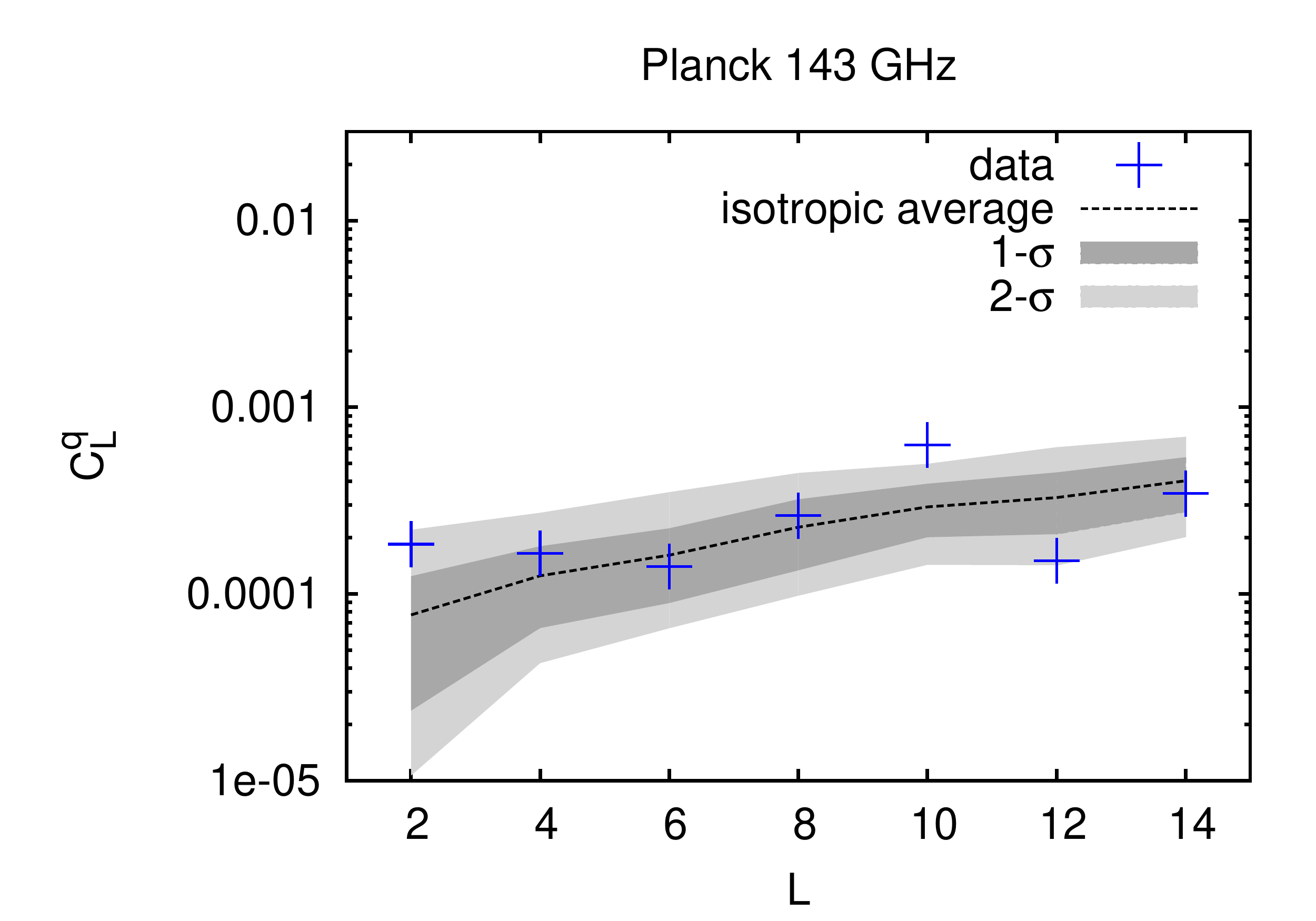}
\includegraphics[width=0.22\columnwidth,angle=-90]{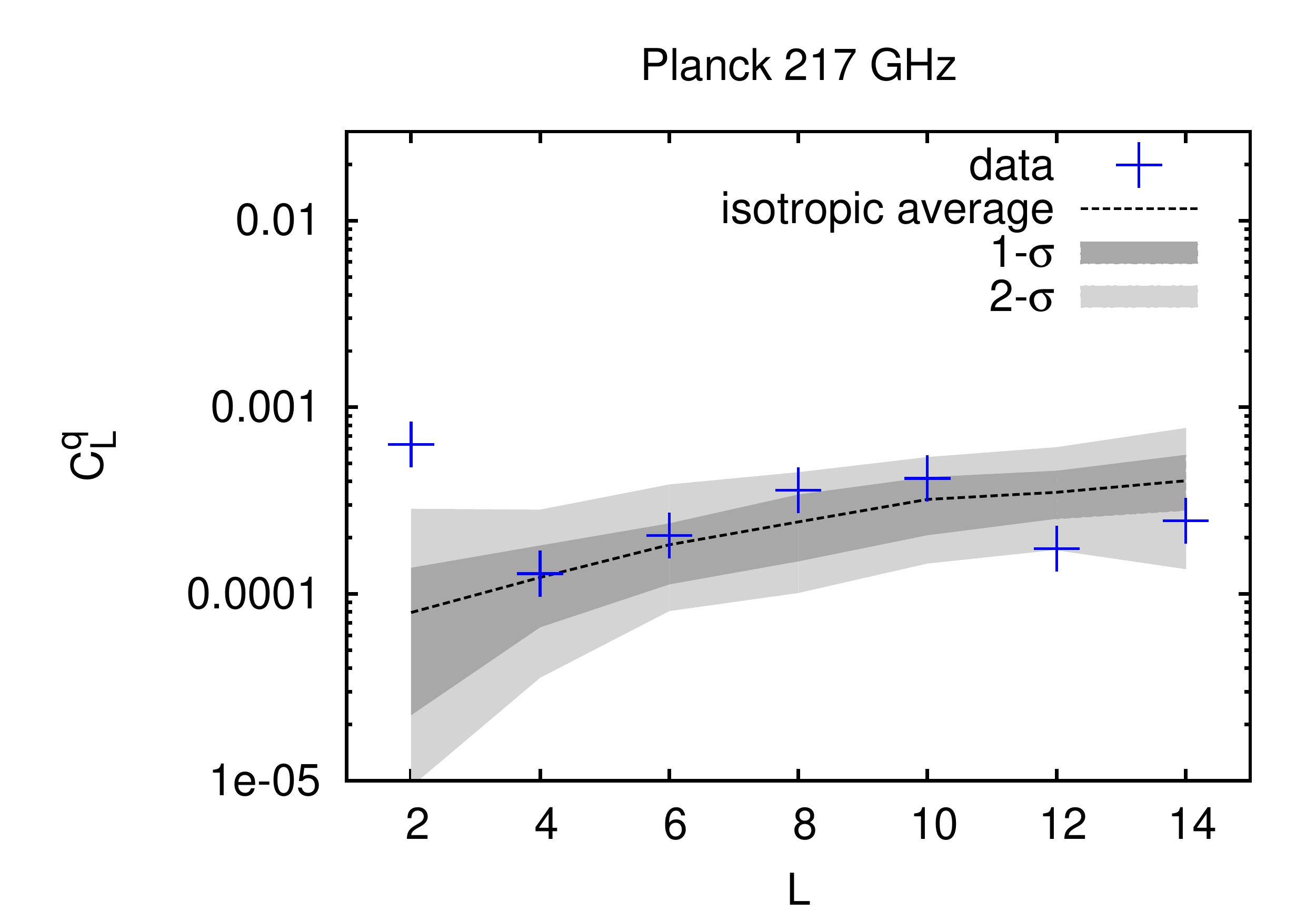}
\includegraphics[width=0.22\columnwidth,angle=-90]{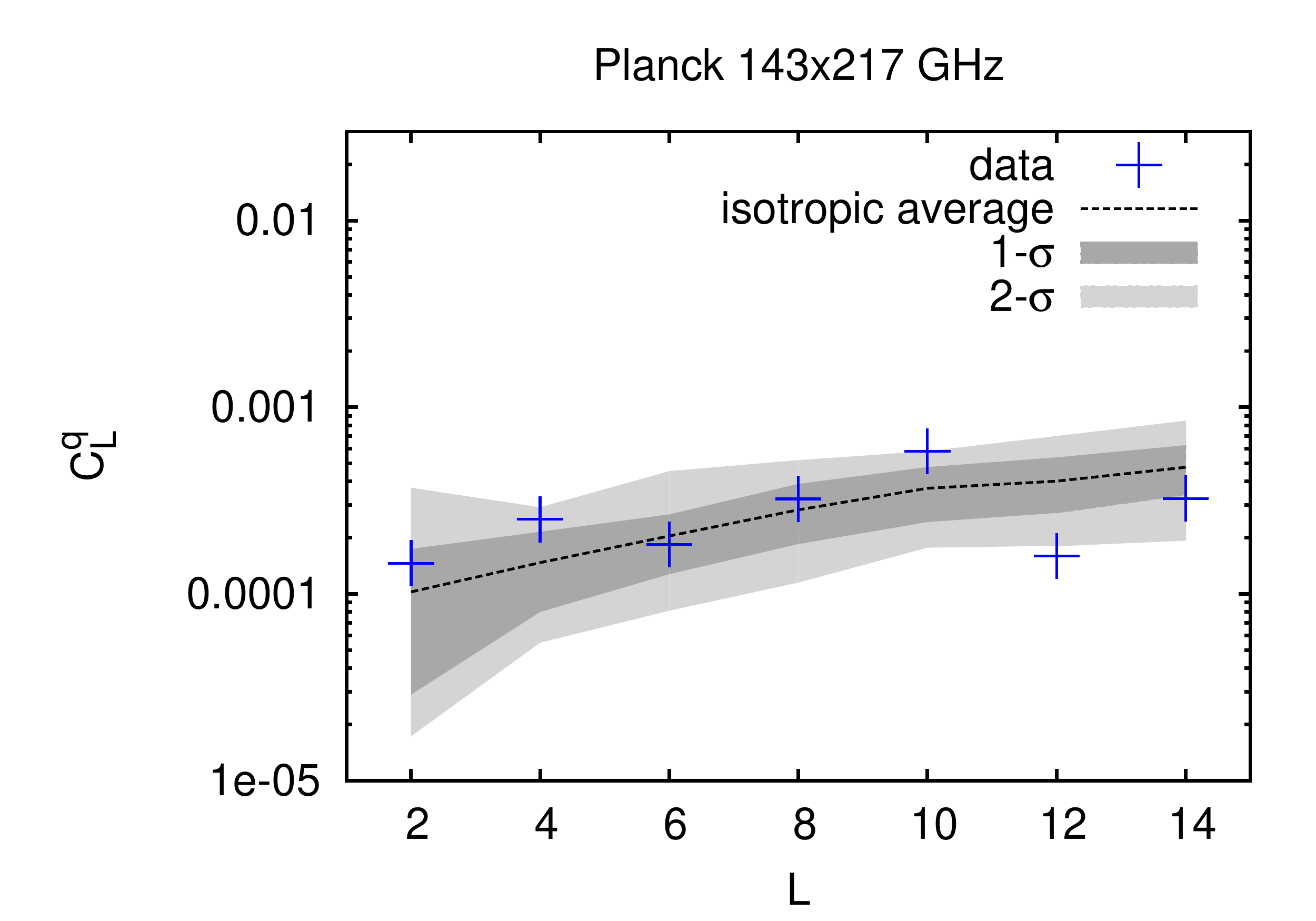}
\end{center}
\caption{$C^{q}_L$ coefficients, as given by Eq.~\eqref{cql}, reconstructed from Planck~2015 data at frequencies $143$\,GHz (left) and $217$\,GHz (centre) as well as their cross-correlation (right). Dark grey and light grey bands correspond to 
68\% C.L. and 95\% C.L. intervals, respectively.}\label{143}
\end{figure}

To assess the sensitivity of the Planck data to SA, we start with the $C^q_L$-statistics defined by 
\begin{equation}
\label{cql}
C^q_L = \frac{1}{2L+1} \sum_M |q_{LM}|^2 \; ,
\end{equation}
(the upperscript 'q' here serves to distinguish the $C^q_L$
coefficients from the amplitudes $C_l$ describing the angular power spectrum). This
has been used in Refs.~\cite{Ramazanov:2013wea, Ramazanov:2012za,
  Rubtsov:2014yua} to constrain models of the (pseudo)Conformal
Universe and anisotropic inflationary scenarios with one vector
field. The reconstructed $C^q_L$ coefficients are shown in
Fig.~\ref{143}. For the sake of completeness, we consider also higher
multipoles of SA up to $L=14$. The data at frequency $143$\,GHz is
consistent with the hypothesis of SI (within $95\%$ C.L.). At the same
time, the data at $217$\,GHz exhibits a quadrupolar SA. We attribute
this departure from SI to unknown systematic effects. These are
typically uncorrelated between different frequency bands, and thus are
mitigated upon cross-correlating the data. It is indeed the case, as
is clearly seen from Fig.~\ref{143}.  A similar observation was made
with the Planck~2013 dataset~\cite{Rubtsov:2014yua}. In that case,
however, the signal of (non-cosmological) SA was identified at
$L=4$. Once again, the consistency with SI was restored in the
cross-correlated data.  In what follows, we stick to the latter as the
one least plagued by possible systematic effects.

 \begin{figure}[tb!]
\begin{center}
\includegraphics[width=0.60\columnwidth,angle=-90]{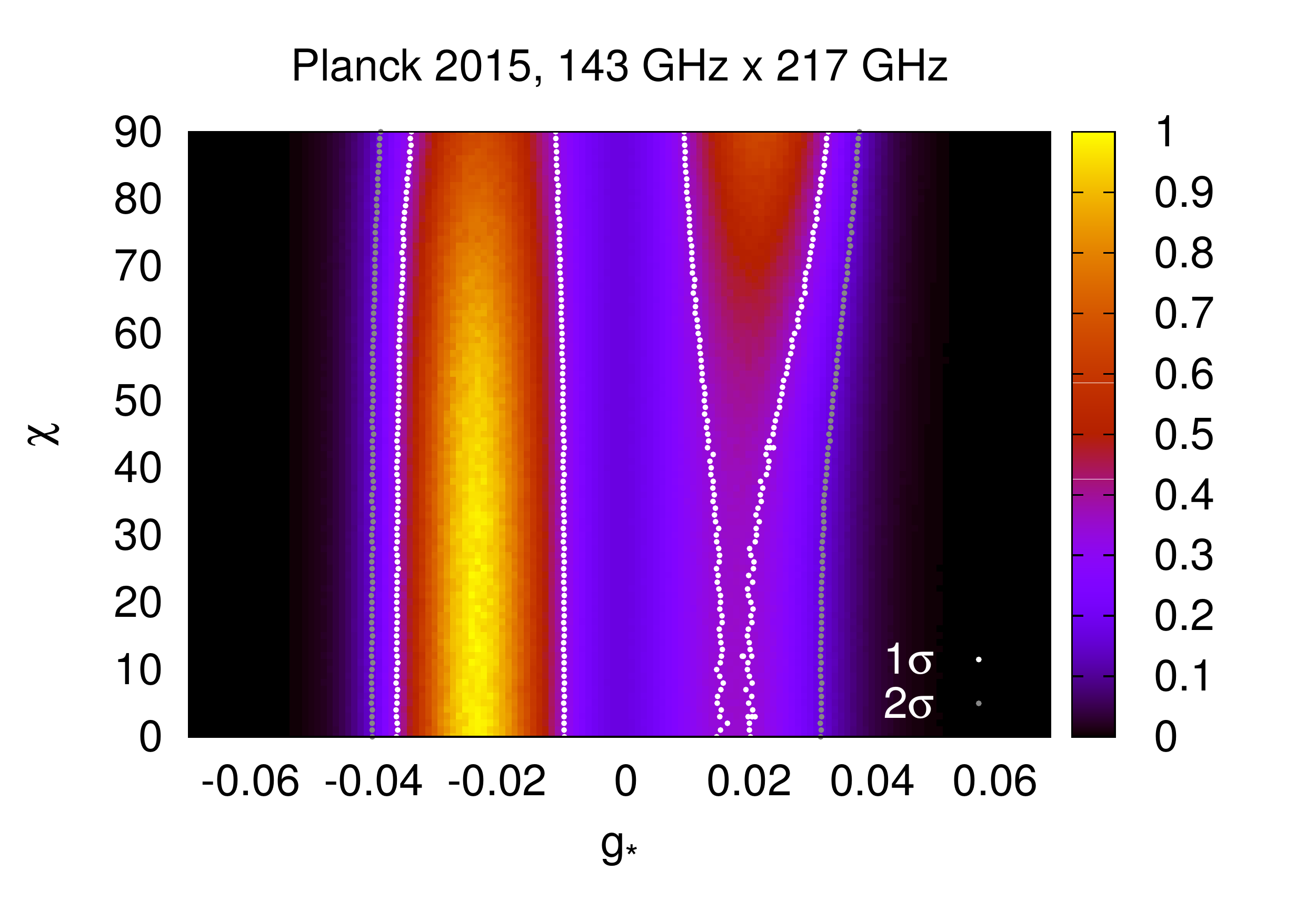}
\caption{The joint likelihood ${\cal L}_J (g_*, \chi)$, obtained from Eq.~\eqref{likelihoodf} by marginalising over the directions ${\bf a}$ and ${\bf c}$, is plotted as a function of the general quadrupole amplitude $g_*$ and shape $\chi$; $68\%$~C.L. and $95\%$~C.L. regions are outlined. The cross-correlation of the $143$\,GHz and $217$\,GHz Planck~2015 maps has been used.}\label{likelihood}
\end{center}
\end{figure}

 \begin{figure}[tb!]
\begin{center}
\includegraphics[width=0.30\columnwidth,angle=-90]{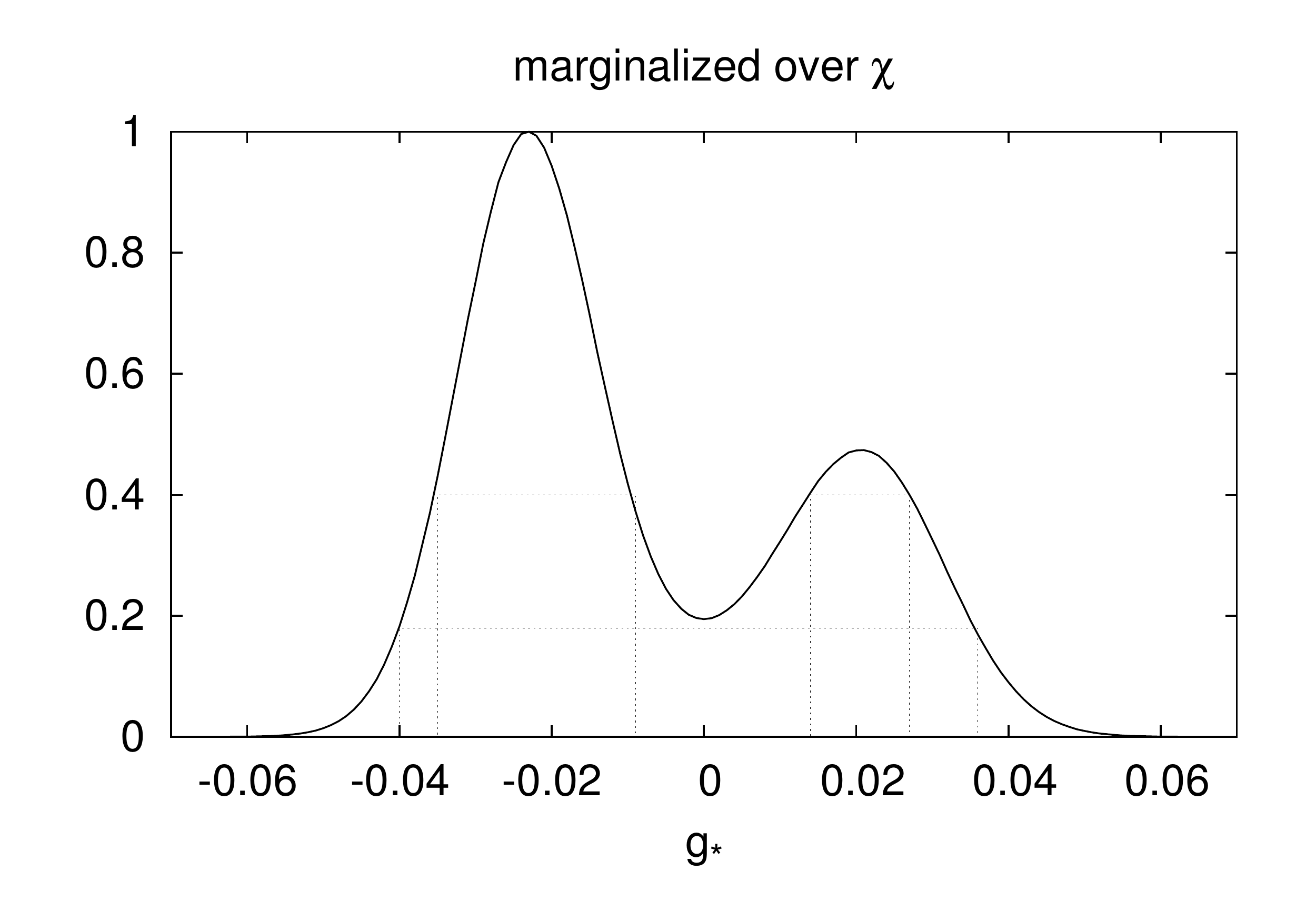}
\includegraphics[width=0.30\columnwidth,angle=-90]{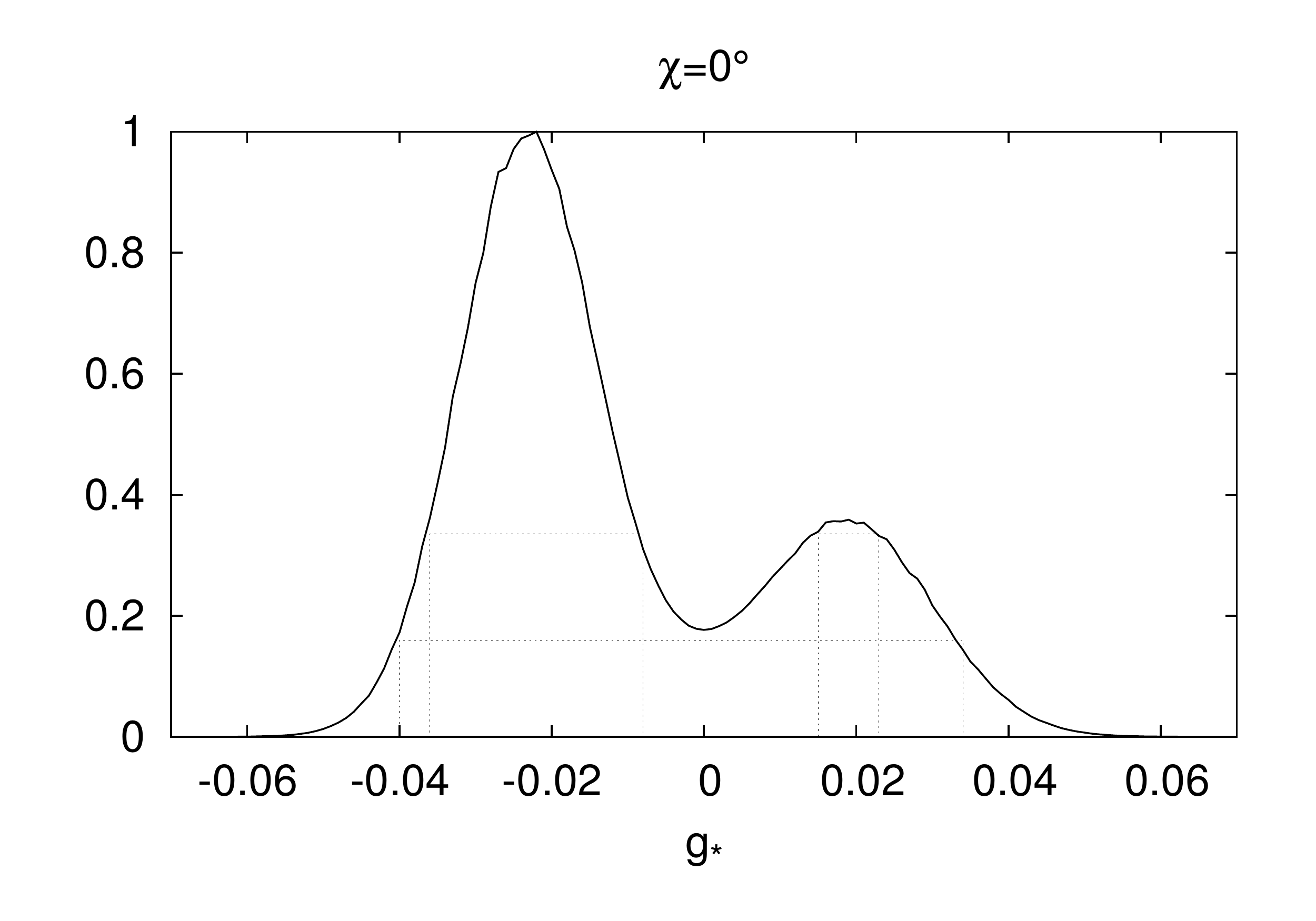}
\includegraphics[width=0.30\columnwidth,angle=-90]{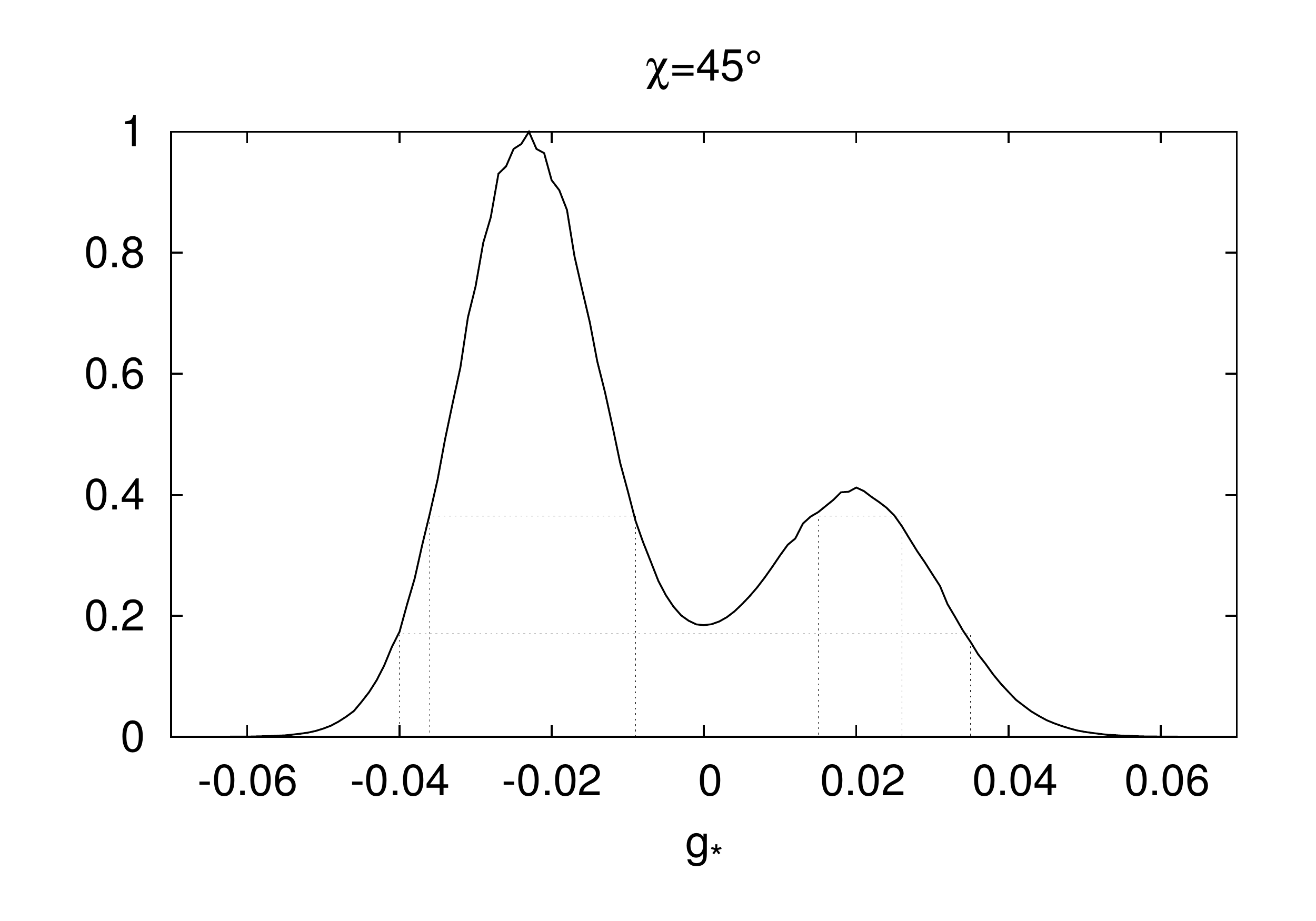}
\includegraphics[width=0.30\columnwidth,angle=-90]{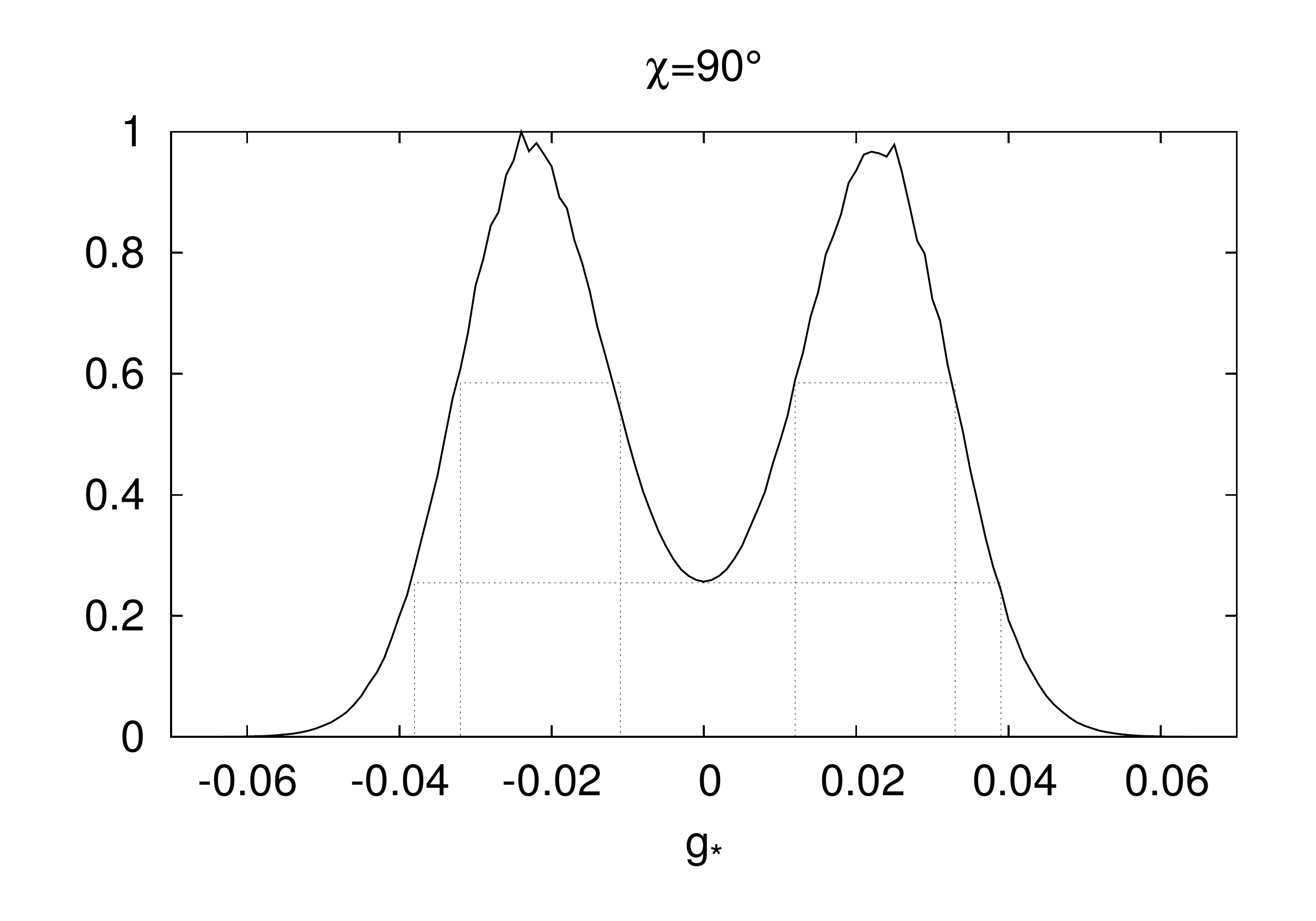}
\end{center}
\caption{Likelihood of the amplitude $g_*$. The top left panel corresponds to the likelihood ${\cal L}_J (g_*, \chi)$ marginalised over the possible values of the angle $\chi$, assuming the latter is homogeneously distributed. The truncation of the joint likelihood ${\cal L}_J (g_*, \chi)$ of Fig.~\ref{likelihood} is shown for $\chi =0^{\circ}$ (top right), $\chi=45^{\circ}$ (bottom left) and $\chi= 90^{\circ}$ (bottow right). Here we used the cross-correlation of the $143$\,GHz and $217$\,GHz Planck~2015 data.}\label{likelihoods}
\end{figure}

In view of our objectives, the $C^q_L$-statistics is not enough, since the $C^q_2$ coefficients are insensitive to the possible $\chi$-dependence of a general quadrupole.\footnote{Indeed, upon substituting Eq.~\eqref{q2mg} into Eq.~\eqref{cql}, 
we see that the parameter $\chi$ drops out.} We find that the strategy adopted in Ref.~\cite{kim} is more appropriate for our purposes here. 
Given that the estimator $\hat q_{LM}$ is affected by a large number
of random quantities, including noise realisation and random
correlations of CMB signal and foregrounds, one may consider a
Gaussian likelihood function,
\begin{equation}
\label{likelihoodf}
{\cal L} (g_*, \chi, {\bf a}, {\bf c})=\frac{1}{\sqrt{2\pi |\mbox{det} {\bf F}^{-1}}|} \mbox{exp} 
\left(-\frac{1}{2}\left[\hat{{\bf q}}-{\bf q} (g_*, \chi, {\bf a}, {\bf c}) \right]^{\dagger}{\bf F}^{-1}
\left[\hat{{\bf q}}- {\bf q} (g_*, \chi, {\bf a}, {\bf c}) \right] \right) \; .
\end{equation}
Substituting Eqs.~\eqref{q2mg} and~\eqref{c2M} into Eq.~\eqref{likelihoodf}, one obtains a posterior distribution for $g_*$, $\chi$, and the vectors ${\bf a}$ and ${\bf c}$. 
We then marginalise over these ${\bf a}$ and ${\bf c}$, and find the joint distribution of $g_*$ and $\chi$: ${\cal L}_J (g_*, \chi)$.  The result is shown in Fig.~\ref{likelihood}. 
This distribution is not Gaussian because the parameters $\chi$, ${\bf a}$, and ${\bf c}$ enter non-linearly in $q_{2M}$, see Eq.~\eqref{q2mg}.

We see from Fig.~\ref{likelihood} that the data treats different values of the parameters $\chi$ non-uniformly. For small $\chi$, when the axial symmetry is restored, 
the data prefers negative $g_*$. For larger $\chi$, the preference shifts towards positive values of $g_*$. Notably, in the limit $\chi \rightarrow 90^{\circ}$, the distribution becomes symmetric 
with respect to a change of the sign of $g_*$. This is especially clear from the distributions of Fig.~\ref{likelihoods} obtained from ${\cal L}_J (g_*, \chi)$ of Fig.~\ref{likelihood} 
by cutting the latter at a given value of the angle $\chi$ (we choose $\chi=0^{\circ}, 45^{\circ}, 90^{\circ}$). In Fig.~\ref{likelihoods} we also plot the likelihood marginalised over possible values of the angle $\chi$. At this level, 
we assume a homogeneous distribution for $\chi$. The final limits on the amplitude $g_*$ are summarised in Table~\ref{Table1}.
% \mik{Notice that the constraint on $g_*$ is only on its absolute value for the case $\chi = 90^{\circ}$, because of the aforementioned symmetrisation.}

The physical reason for the symmetrisation of the likelihood at large values of the angle $\chi$ is as follows. In the case $\chi = 90^{\circ}$, Eq.~\eqref{genparametr} reduces to
\begin{equation}
\nonumber
\sum_M q_{2M} Y_{2M} (\hat{{\bf k}}) \propto g_* \left( Y_{2,1} (\tilde{\vartheta}, \tilde{\varphi})-Y_{2,-1} (\tilde{\vartheta}, \tilde{\varphi})\right) \; ,
\end{equation}
where we omitted an irrelevant constant factor. The r.h.s.\ here is symmetric under the coordinate transformation $\tilde{\varphi} \rightarrow \tilde{\varphi}+ \pi$, supplemented 
by a change of sign for the amplitude $g_*$, i.e., ${\cal L} (g_*, \chi =90^{\circ}, {\bf a}, {\bf c})={\cal L} (-g_*, \chi=90^{\circ}, -{\bf a}, {\bf c})$. 
Therefore, upon marginalising over the directions of the vectors ${\bf a}$ and ${\bf c}$, we get ${\cal L}_J (g_*, \chi =90^{\circ})={\cal L}_J (-g_*, \chi=90^{\circ})$. The two cases are therefore physically indistinguishable. %This explains the symmetrisation. 

\begin{table}[htb!]
\begin{tabular}{|c|c|c|}
\hline 
$\chi$ &  $g_*$, $68\%$ C.L. limit &  $g_*$, $95\%$ C.L. limit  \\
\hline
$\chi=0^{\circ}$ &  $-0.037 <g_* <-0.008$~~~$0.014<g_*<0.023$ &   $-0.041 <g_* <0.034$  \\
\hline
$\chi= 45^{\circ}$  & $-0.037 <g_* <-0.009$~~~$0.014<g_*<0.026$ & $-0.041 <g_* <0.035$ \\
\hline
$\chi= 90^{\circ}$  &$0.011<|g_*| <0.033$ & $|g_* |<0.039$\\
\hline 
Arb. $\chi$  &  $-0.036 <g_* <-0.009$~~~$0.013<g_*<0.027$ & $-0.041 <g_* <0.036$\\
\hline
Arb. $\chi$, Planck~2013 & $-0.015<g_*<0.016$  & $-0.028<g_*<0.030$\\
\hline 
\end{tabular}
\caption{\footnotesize{Planck~2015 68\% and 95\% C.L. limits on the amplitude $g_*$ of the
 quadrupole for different choices of the parameter $\chi$. For the sake of comparison, here we also show Planck~2013 limits on the amplitude of the axisymmetric quadrupole ($\chi=0^{\circ}$). In all cases the cross-correlated data has been used.}}\label{Table1}
\end{table}

%\begin{table}[htb!]
%\begin{tabular}{|c|c|c|c|c|c|}
%\hline 
%$\chi$ &  $\chi =0$ & $\chi=\frac{\pi}{4}$ & $\chi=\frac{\pi}{2}$ & Arb. $\chi$ & $\chi =0$ (2013)  \\
%\hline
%$g_*$, $95\%$ &  $0.001 \pm 0.042$ & $-0.002 \pm 0.039$ & $ 0.000 \pm 0.041$ & $-0.002 \pm 0.039$ & -  \\
%\hline
%$g_*$, $68\%$ & -  &-  &-  &- &-\\
%\hline
%\end{tabular}
%\caption{\footnotesize{Planck 68\% and 95\% C.L. limits on the amplitude $g_*$ of the
% quadrupole for different choices of the parameter $\chi$. For the sake of comparison, here we also show Planck~2013 limits on the amplitude of the axisymmetric quadrupole ($\chi=0$).}}\label{Table1}
%\end{table}

 \begin{figure}[tb!]
\begin{center}
\includegraphics[width=0.30\columnwidth,angle=-90]{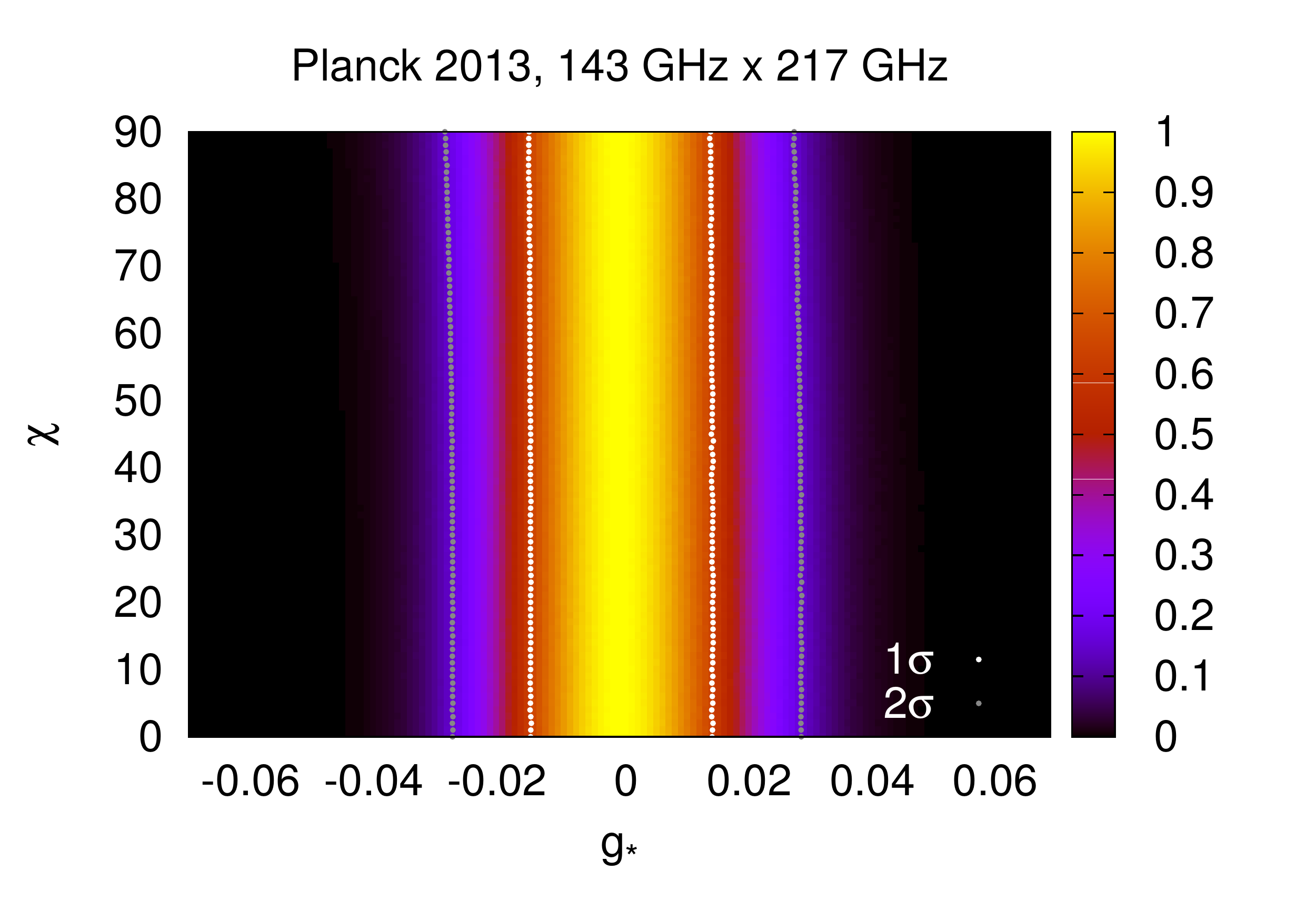}
\includegraphics[width=0.30\columnwidth,angle=-90]{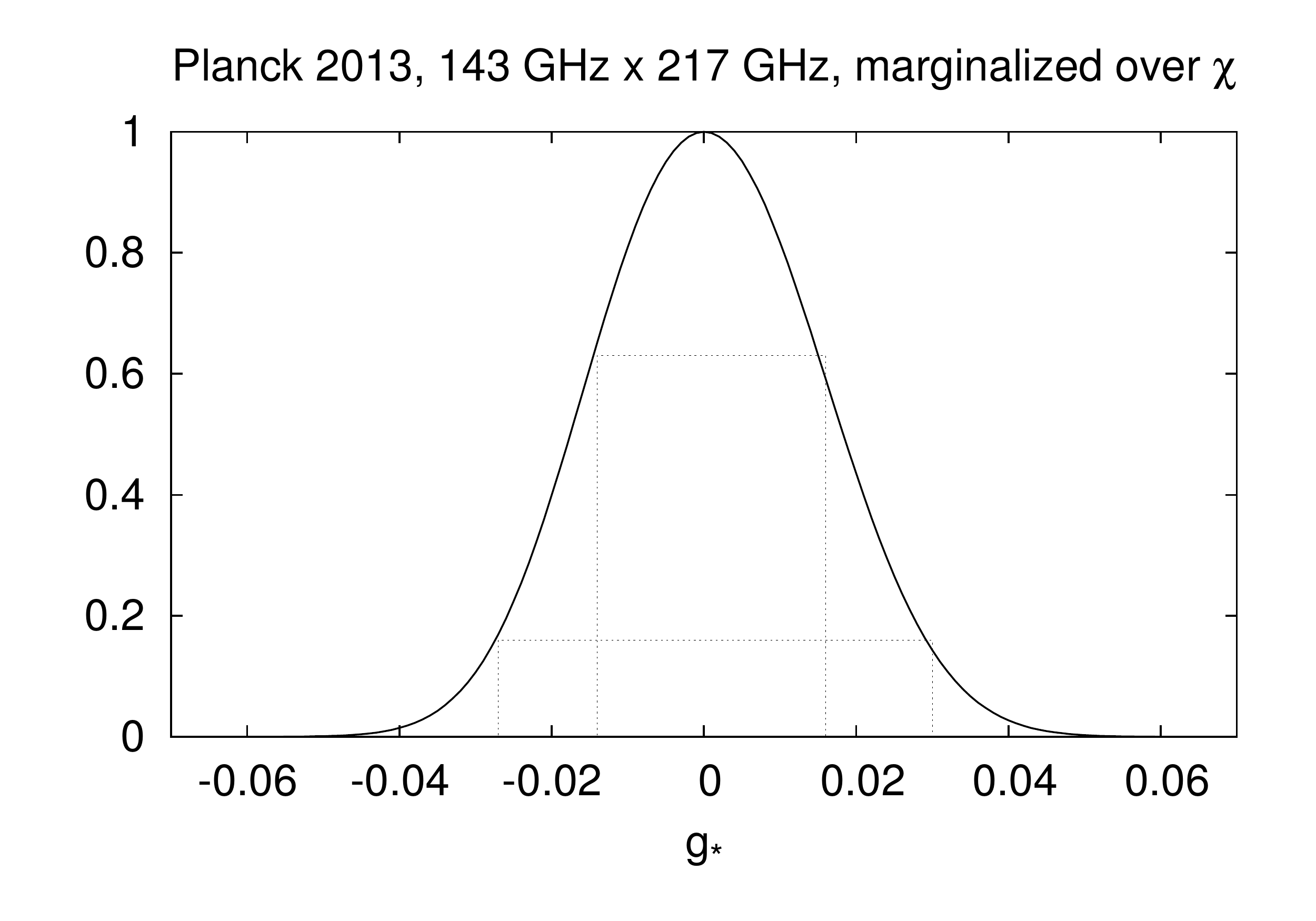}
\end{center}
\caption{Left panel: likelihood of the parameters $g_*$ and $\chi$ derived from Planck~2013 cross-correlated data. Right panel: the same, but now marginalised over possible 
values of $\chi$. On both plots $68\%$ C.L. and $95\%$ C.L. regions are outlined.}\label{likelihoods2013}
\end{figure}

Now let us compare the Planck~2013 and~2015 datasets. The joint
likelihood ${\cal L}_J (g_*, \chi)$ obtained from Planck~2013 data is
shown in Fig.~\ref{likelihoods2013}. Notably, this distribution corresponds to nearly zero best-fit value of the amplitude $g_*$. The reason is that the coefficients $h_{LM}$ reconstructed from
Planck~2013 maps are very close (i.e., well within $1\sigma$ interval) to the average
value obtained from statistically isotropic Monte-Carlo maps. As a result, the distribution ${\cal L}_J (g_*, \chi)$ is independent on the angle $\chi$---evidently, for vanishing SA the data can not discriminate between 
different quadrupole shapes. Our final limits on the amplitude $g_*$ are shown in Table~\ref{Table1}. They agree very well with the limits deduced in Ref.~\cite{kim} for the case of the axisymmetric quadrupole.

%\mik{QUESTION: Is it possible to
%  comment briefly on why this is the case?} 
% That
%conclusion matches very well the results of
%Refs.~\cite{Thorsrud:2013kya, Thorsrud:2013mma}. \mik{THAT LOOKS LIKE
%  WRONG REFS!!!} As a by-product, the likelihood of the amplitude
%$g_*$ shown ibid has a nearly Gaussian form, which translates into the
%stronger limits on SA given in Table~\ref{Table1}.

%%%%%%%%%%%%%%%%%%%%%%%%%%%%%%%%%%%%%%%%%%%%%%%%%%%%%%%%%%%%%%%%%%%%%%%%%%%%%%

%%%%%%%%%%%%%%%%%%%%%%%%%%%%%%%%%%%%%%%
\section{Anisotropic inflation with multiple vector fields}\label{ssec:vec}
%%%%%%%%%%%%%%%%%%%%%%%%%%%%%%%%%%%%%%%
Typically, in early Universe scenarios, the amplitude 
$g_*$ and the angle $\chi$ are random variables with statistical
properties determined by the intrinsic model parameters. That is, there is no one-to-one correspondence between the quadrupolar amplitude and shape and the theoretical parameters in the Lagrangian. 
In this situation, the relevant constraints can not be immediately inferred from those of Table~\ref{Table1}. We perform such analysis in this Section for inflationary scenarios in which several Maxwellian spectator fields are non-minimally coupled to an inflaton~\cite{Watanabe:2009ct, Watanabe2, Bartolo:2012sd, Lyth:2013sha, Thorsrud:2013kya, Thorsrud:2013mma}. A similar analysis for the case of the (pseudo)Conformal Universe (but with Planck~2013 data) can be found in Ref.~\cite{Rubtsov:2014yua}: we briefly revisit those limits with the new dataset at the end of the Section. 

%\subsection{Basics} 
One way to achieve SA in inflation is, by
virtue of their directional nature, to introduce vector fields.  The most well-known example is
the model where the Maxwellian fields are coupled to the inflaton
itself~\cite{Watanabe:2009ct, Watanabe2}. In that case the vectors' $U(1)$
gauge invariance is preserved, and one has a chance to achieve SA without
developing catastrophic ghost instabilities~\cite{ghost}. In the
literature, the scenario with a single gauge field is the most
popular one. In this case, one deals with a directional dependence in the
scalar perturbations power spectrum which is axisymmetric and
characterised by a \emph{negative} amplitude
$g_*$~\cite{Ohashi:2013qba}. In the setup with multiple gauge
  fields~\cite{Thorsrud:2013kya, Thorsrud:2013mma} the axial symmetry
  is broken and the resulting SA is a general quadrupole.

We consider the gauge sector action:  
\begin{equation}
S_{\cal A} =-\frac{1}{4}\int d^4 x \sqrt{-g} \cdot f^2 (\phi) \cdot \sum_{a=1}^n F^{\mu \nu}_a F_{\mu \nu a} \; .
\label{action}
\end{equation}
Here the subscript $a$ runs over the collection of $n$ gauge fields $A^a_\mu$ for which $F^a_{\mu\nu}$ is the strength. For simplicity we assumed that the coupling of the gauge 
fields to the inflaton is universal, but generalisations are straightforward and do not significantly change our results. If we were to take a trivial kinetic gauge function, $f(\phi)=1$, the contribution of each gauge field would redshift away adiabatically due to the expansion of the universe. This decay can be prevented by choosing appropriate dynamics for $f(\phi)$. 

One well justified possibility is $f(\phi) \propto a^{-2}$~\cite{Watanabe:2009ct}; this might in fact be the only sensible option since it is a quite generic attractor of this system~\cite{Watanabe:2009ct, Yamamoto:2012tq, Kanno:2010nr, Dimopoulos:2010xq, Hervik:2011xm, Thorsrud:2012mu, Ito:2015sxj}. In that case,    
the contribution of the ``electric'' energy density remains constant during inflation\footnote{The opposite choice 
$f(\phi) \propto a^2$---also an attractor for an inverted coupling function $f(\phi)\rightarrow f^{-1}(\phi)$---would result into 
a ``magnetic'' energy density which is constant with time; this option is
typically invoked in the context of
magnetogenesis~\cite{Subramanian:2015lua}.}. This makes it possible to generate non-trivial SA described by~\cite{Watanabe:2009ct, Watanabe2, Bartolo:2012sd, Thorsrud:2013kya, Thorsrud:2013mma} 
\begin{equation}
q_{2M}  = \sum_a g^a_* \int (\hat{\mathbf{k}} \cdot \hat{{\bf E}}^a_{cl})^2 Y^{*}_{2M}(\hat{\mathbf{k}}) d\Omega = \sum_a\frac{8\pi g^a_*}{15} Y^{*}_{2M} (\hat{{\bf E}}^a_{cl})  \; .
\label{q2mmult}
\end{equation}
Here ${\bf E}^a_{cl}$, is the classical ``electric'' field. See the discussion below on its origin. The amplitudes $g_*^a$ are given by 
\begin{equation}
g^a_* (k) =-\frac{24}{\epsilon} \cdot \frac{\left({\bf E}^a_{cl}(\tau_0) \right)^2}{V(\phi_k)} \cdot N^2_k \; ,
\label{gstar}
\end{equation}
where $N_k$ is the number of e-folds between horizon crossing of mode $k$ and the end of inflation; $\epsilon$ is the standard slow roll parameter and $V (\phi)$ is the inflaton potential. The amplitudes $g^a_*$ here are not to be 
confused with the amplitude $g_*$ of the general quadrupole defined from Eq.~\eqref{genparametr}. While the former are always negative, the latter is allowed to 
take on positive values as well. 

%\begin{table}[htb!]
%\hspace{-0.4cm}
%\begin{tabular}{|l|c|c|c|c|}
%\hline 
%& WMAP9/V & Planck/143 & Planck/217& Planck/143$\times$217 \\
%\hline
%Axisymmetric%\footnote{This should be rather viewed as the nominal constraint 
%commonly assumed in literature from the direct comparison with the anomalously quadrupole observed in the $W$ band 
%of the WMAP 5- and 7-year releases} 
%&$|g_*| <0.072$&- & -& $|g_*|<0.026$\\
%\hline
%Inflation, $n=1$ &$N <82 \left(\frac{60}{N_{k_{min}}} \right)^2$  & - & - &  $N_\text{xtr} <35 \left(\frac{60}{N_{k_{min}}} \right)^2$ \\
%\hline
%Inflation, $n=3$ &- &- & -& -\\
%\hline 
%Inflation, $n=10$ &- & - & -& -\\
%\hline 
%Inflation, $n=100$ &- &- &- &-\\
%\hline
%\end{tabular}
%\caption{\footnotesize{Planck $95\%$ C.L. constraints on the parameters of anisotropic models from the non-observation 
%of statistical anisotropy in the CMB sky. These include anisotropic inflation, conformal rolling scenarios and galileon genesis. \fede{[Add solid inflation]}}}\label{constrearly}
%\end{table}

The ``electric'' fields ${\bf E}^a_{cl}$ have two sources. 
One is purely classical, corresponding to the attractor solution of the background equations of motion~\cite{Watanabe:2009ct}. 
The second is 
due to quantum fluctuations which get enhanced and stretched during inflation (and finally classicalise once they leave the horizon)~\cite{Bartolo:2012sd}. 
This is an infrared component which is built up of all modes, processed by inflation, which are now beyond our observable horizon. One thus writes ${\bf E}^a_{cl}$ as follows
\begin{equation}
\nonumber 
{\bf E}^a_{cl} (\tau_0)={\bf E}^a_0 + {\bf E}^a_{IR} (\tau_0) \; .
\end{equation}
 Here ${\bf E}^a_{IR}(\tau_0)$ is the status of the infrared vector at 
the time when the gauge mode matching the size of our observable universe crossed the horizon, $\tau_0=-1/\mathcal{H}_0$; here ${\cal H}_0 \sim k_{min}$ denotes the 
present Hubble rate in conformal time units. Modes crossing the horizon after $\tau_0$ are ignored because they look inhomogeneous from our point of view and hence 
do not contribute significantly to the global asymmetry parametrised by $g_*$. See Ref.~\cite{Thorsrud:2013mma} for a careful discussion of this point. We model the vector as a Gaussian field with zero mean and variance~\cite{Bartolo:2012sd} 
\begin{equation}
\label{elcl}
\langle \left( {\bf E}^a_{IR} (\tau_0) \right)^2\rangle =\frac{9H^4 (k_{min})}{2\pi^2} N^e \; ,
\end{equation}
where $N^e=N_{tot}-N_{k_{min}} \sim N_{tot}-N_{{\cal H}_0}$ is the ``extra'' e-folds of inflation, namely, the number of e-folds between start of inflation and $\tau_0$.   

%\subsubsection{Constraints on number $N_\text{xtr}$}

As far as the constraining procedure is concerned, 
barring cancellations, our limits would be conservative, as we attribute all anisotropy to one component only; 
any other additional anisotropy would only exacerbate the discrepancy with the data.  
Notice furthermore that in multivector scenarios background isotropy is attractive, and the 
anisotropy is then expected to only arise due to the infrared fluctuations. Hence, we will be interested in the case for which the purely classical ``electric'' fields ${\bf E}^a_0$ give 
a negligible contribution to SA, i.e., ${\bf E}^a_{cl} \approx {\bf E}^a_{IR}$. Then, the ``extra'' number of e-folds $N^e$ 
is the only relevant parameter which affects SI and generates SA.

Before diving into the routine of the constraining procedure, let us
make a short remark. By glancing at Eq.~\eqref{gstar}, one may see that the amplitude $g_*$ is scale-dependent.  Despite this dependence being relatively mild, it may essentially bias our constraints, because Planck has access to a wide range of multipoles. In reality, the dependence present in $N_k$ and the one due to the slow roll potential $V(\phi_k)$ compensate each other with a high accuracy:
\begin{equation}
\nonumber
\frac{\partial \ln | g^a_* | }{\partial  \ln k  } \approx  -\frac{\partial V(\phi_k)}{\partial \ln k} -\frac{2}{N_{k}} \approx -(n_s-1) -\frac{2}{N_k}  \; .
\end{equation}
Substituting the experimental central value $n_s-1 \approx -0.04$ and $N_k \approx 60$, we observe that two values 
on the r.h.s.\ approximately sum up to zero. Thus, we can safely set $k=k_{min}$ in Eq.~\eqref{gstar}.

 \begin{figure}[tb!]
\begin{center}
\includegraphics[width=0.60\columnwidth,angle=-90]{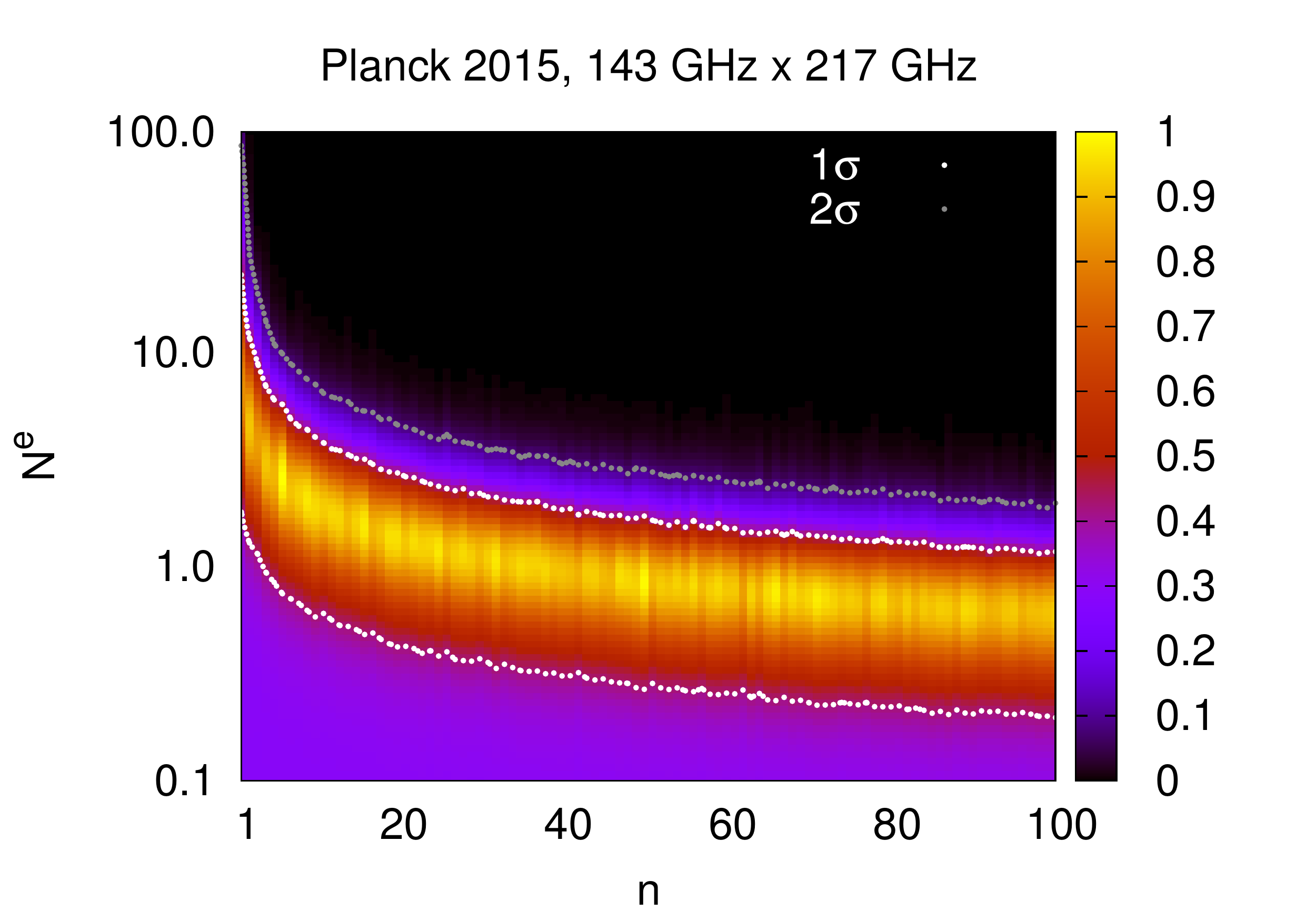}
\end{center}
\caption{Distribution of the ``extra'' number of e-folds in inflation, $N^e$, as the function of the number $n$ of vector fields; $68\%$~C.L. and $95\%$~C.L. regions are outlined.}\label{L_nN}
\end{figure}

It is convenient to rewrite the amplitudes $g^a_*$ given by Eq.~\eqref{gstar} as follows 
\begin{equation}
\nonumber 
g^a_* =-({\bf e}^a)^2 \; ,
\end{equation}
where ${\bf e}^a$ are Gaussian vectors collinear with ${\bf E}^a_{IR}$, characterised by zero means and variances 
\begin{equation}\label{gauzz}
\langle (e^a_i)^2 \rangle =96 N^e N^2_{k_{min}}  {\cal P}_{\zeta} (k_{min}) \; .
\end{equation}
%\begin{equation}\label{gauzz}
%\langle (a^a_i)^2 \rangle =96 N^2_{CMB} (k_*) N(k_*) {\cal P}_{\zeta} (k_*) \; .
%\end{equation}
Note that each component ($i=\{x,y,z\}$) has the same variance since the background spacetime is rotationally symmetric to very good accuracy. 
Here we made use of the slow roll relation 
\begin{equation}
\nonumber 
{\cal P}_{\zeta} (k)=\frac{3H^4}{8\pi^2 V(\phi) \epsilon} \; , 
\end{equation}
in order to get rid of the dependence on the potential $V(\phi)$ and
the slow-roll parameter $\epsilon$.  

To constrain the number $N^e$ we
use the following strategy: starting from a given value of $N^e$ we
generate an ensemble of $10^4$ sets of Gaussian vectors
${\bf e}^a$ from Eq.~(\ref{gauzz}). For each set, we
calculate the coefficients $q_{2M}$ defined by the
relation~\eqref{q2mmult} and compute the likelihood of Eq.~\eqref{likelihoodf} using Planck data. The latter is then averaged over the ensemble of sets. We run this algorithm for different numbers
of Maxwellian fields: $n=[1,100]$ in steps of~1 and different $N^e$
values. The results are presented in Fig.~\ref{L_nN}. Notice that the best fit
value of $N^e$ falls roughly as $1/\sqrt{n}$ with the number of
fields, in accordance with theoretical
expectations~\cite{Thorsrud:2013kya, Thorsrud:2013mma}. 

For comparison purposes, the constraint on the number of e-folds $N^e$ in the case of the single vector field read: 
\begin{equation}\label{68cl}
N^e <71 \cdot \left(\frac{60}{N_{k_{min}}} \right)^2 \qquad 95\%~\mbox{C.L.}\; ,
\end{equation}
($ 2 <N^e <18 $ at $68\%$ CL for $N_{k_{min}}=60$). This limit is only a very moderate improvement compared to the analogous WMAP9 limit of Ref.~\cite{Ramazanov:2013wea}.

\paragraph{The (pseudo)Conformal Universe revisited.}
 
Before closing, let us comment on some immediate implications of Eq.~\eqref{68cl} for the (pseudo)Conformal Universe~\cite{Rubtsov:2014yua}. The latter is an alternative to inflation, which attributes the approximate flatness of primordial peturbations to the assumed conformal symmetry of the early Universe~\cite{Hinterbichler:2011qk, Rubakov:2009np, Creminelli:2010ba}. 
As far as SA is concerned, the (pseudo)Conformal Universe is practically equivalent (modulo a constant factor) to inflation augmented with a single Maxwellian 
field~\cite{Libanov:2010nk, Ramazanov:2013wea}\footnote{In the case of the (pseudo)Conformal 
Universe, there is an additional contribution to SA, which is a general quadrupole~\cite{Libanov:2010nk}. This, however, has an amplitude decreasing as $g_* \propto k^{-1}$ with the wavenumber $k$. Hence, it makes a negligible imprint on the CMB.}. 
This is true at least for the subclass of (pseudo)Conformal Universe models in which the cosmological perturbations of interest remain frozen after the conformal phase and before
 the hot stage/reheating~\cite{Hinterbichler:2011qk, Creminelli:2012qr, Libanov:2015iwa, Rubakov:2009np, Libanov:2010nk, Libanov:2011bk, Creminelli:2010ba} ('sub-scenario A' in the notation of Ref.~\cite{Rubtsov:2014yua}). 
The constraint of Eq.~\eqref{68cl} translates to
\begin{equation}
\label{limitconformal}
h^2 \ln \frac{H_0}{\Lambda} <1.0 \qquad 95\%~\mbox{C.L.} \; .
\end{equation}
Here $h^2$ is the parameter which governs the non-trivial evolution in 
the (pseudo)Conformal Universe; $H_0$ is the Hubble rate today, and $\Lambda$ is the cutoff on the modes feeding into SA, see Ref.~\cite{Rubtsov:2014yua} for details\footnote{Note that the limit~\eqref{limitconformal} does not apply to the situation in which the conformal phase and the hot epoch are separated by a long intermediate stage ('sub-scenario B' in the notation of Ref.~\cite{Rubtsov:2014yua}), where cosmological perturbations follow a non-trivial evolution. This case involves all multipoles of SA~\cite{Libanov:2011hh} and requires a special analysis. It was addressed in Ref.~\cite{Rubtsov:2014yua} with Planck~2013 data, and we do not revisit it here.}.

\section*{Acknowledgments} 
%%%%%%%%%%%%%%%%%%%%%%%%%%%%%%%%%%%%%%%%%%%%%%%%%%%%%%%%%%%%%%%%%%%%%%%%%%%%%%

The comparison of Planck data with the predictions of multivector
inflation models is supported by the Russian Science Foundation grant
No. 14-12-01430 (G.R.). The analysis of the (pseudo)Conformal
Universe models is supported by the Russian Science Foundation grant
14-22-00161 (G.R.). F.U. is supported by the ERC grant PUT808 and the
ERDF CoE program.

%%%%%%%%%%%%%%%%%%%%%%%%%%%%%%%%%%%%%%%%%%%%%%%%%%%%%%%%%%%%%%%%%%%%%%%%%%%%%%
\section*{Appendix. Multipole vector representation of a general quadrupole \label{appA}}
%%%%%%%%%%%%%%%%%%%%%%%%%%%%%%%%%%%%%%%%%%%%%%%%%%%%%%%%%%%%%%%%%%%%%%%%%%%%%%

Instead of the quantities $g_*$, $\chi$, ${\bf a}$ and ${\bf c}$, one may wish to 
work in terms of the multipole vectors. In order to do so, we follow Ref.~\cite{Copi:2003kt} and write down the multipole vector definition for a 
quadrupole:
\begin{equation}
\label{mv}
\sum_{M} q_{2M} Y_{2M} = 
A^{(2)} \sum^{1}_{\tilde{M}=-1} \sum^{1}_{J=-1} v^{(2,1)}_J v^{(2,2)}_{\tilde{M}} Y_{1,J} 
Y_{1, \tilde{M}} +C\; ,
\end{equation}
where $v^{(2,1)}$ and $v^{(2,2)}$ are two multipole vectors assumed to be normalised as 
\begin{equation}
\nonumber 
v_{-1} =\frac{1}{\sqrt{2}} \left(v_x +iv_y \right), \qquad v_0 =v_z, \qquad v_1 =-\frac{1}{\sqrt{2}} 
(v_x -i v_y) \; .
\end{equation}
The constant $C$ can be obtained by integrating the left and the 
right hand sides of Eq.~\eqref{mv} over the directions 
of the cosmological mode ${\bf k}$. The result reads
\begin{equation}
\nonumber 
C=-\frac{1}{4\pi} ({\bf v}^{(2,1)} \cdot {\bf v}^{(2,2)})\; .
\end{equation}

Notice that the relation~\eqref{mv} does not fix the signs 
of the multipole vectors and of the amplitude $A^{(2)}$. 
We partially eliminate this ambiguity by requiring, with no loss of generality,
\begin{equation}
\label{signcond}
{\bf v}^{(2,1)} \cdot {\bf v}^{(2,2)} >0 \; .
\end{equation}
The remaining freedom---which is due to the simultaneous change of sign 
of two multipole vectors---does not affect any physical quantity, 
since the vectors always enter only in bilinear combinations. 
To obtain the general quadrupole in the form~\eqref{genparametr} we choose a coordinate system with the $z$ axis aligned 
with one of the multipole vectors. Clearly, we can organise this in two possible ways 
accordingly to the number of multipole vectors. For concreteness, 
we pick the vector $v^{(2,2)}_J$, i.e., $v^{(2,1)}_0= 1$ and $v^{(2,2)}_{\pm 1}=0$. From the 
condition~\eqref{signcond}, it follows that $v^{(2,1)}_z>0$, and 
\begin{equation}
\label{fixcoord}
\sum_{M} q_{2M} Y_{2M}= A^{(2)}\sum^{1}_{J=-1}  v^{(2,1)}_{J} Y_{1,0} 
Y_{1, J}  \; ,
\end{equation}
Finally, we can rotate the coordinate system in such a way 
that the vector ${\bf v}^{(2,1)}$ lies in the $O\tilde{x}\tilde{z}$ plane, 
i.e., $v^{(2,1)}_{-1}=-v^{(2,1)}_{1} = v^{(2,1)}_{x} /\sqrt{2}$. 
To fix the coordinate system completely, we require that $v^{(2,1)}_x>0$. From Eq.~\eqref{fixcoord}, 
we then obtain the quadrupole term in the form~\eqref{genparametr},
\begin{equation}
\label{genmv}
\sum_M q_{2M} Y_{2M} =
A^{(2)} \left\{v^{(2,1)}_{z} Y_{20} \frac{1}{\sqrt{5\pi}} - v^{(2,1)}_{x}\sqrt{\frac{3}{40\pi}} \left[Y_{21}-Y_{2,-1} \right] 
 \right\} \; .
\end{equation}
Now, we observe that there is a one-to-one correspondence between this parametrisation 
(with the coordinate system fixed as discussed above) and that of Eq.~\eqref{genparametr} 
in the region $-\infty<g_*<+\infty$ and $0^{\circ} \leq \chi \leq 90^{\circ}$. 
To make it even clearer, 
we can obtain the explicit relations between $g_*$ and $\chi$ and the 
multipole vectors as 
\begin{equation}
\label{chimv} 
\chi =\arctan \frac{\sqrt{3} v^{(2,1)}_{x}}{2v^{(2,1)}_{z}} =
\arctan \frac{\sqrt{3[1-({\bf v}^{(2,1)} \cdot {\bf v}^{(2,2)})^2]}}{2({\bf v}^{(2,1)} \cdot {\bf v}^{(2,2)})} \; ,
\end{equation}
\begin{equation}
\nonumber
{\bf c} ={\bf v}^{(2,2)} \; , \qquad {\bf a} =
\frac{{\bf v}^{(2,2)} \times \left({\bf v}^{(2,1)} \times {\bf v}^{(2,2)}\right)}
{\left|{\bf v}^{(2,2)} \times \left({\bf v}^{(2,1)} \times {\bf v}^{(2,2)}\right)\right|} \; ,
\end{equation}
and 
\begin{equation}
\label{gstarmv}
g_*=\frac{3A^{(2)}}{8\pi} \sqrt{ ({\bf v}^{(2,1)} \cdot {\bf v}^{(2,2)})^2+3} \; .
\end{equation}
One final remark is in order here. As it follows from 
the discussion above, there are three more coordinate system 
where the quadrupole takes the form~\eqref{genparametr}. One is associated 
with the interchange of the multipole vectors ${\bf v}^{(2,1)} \leftrightarrow {\bf v}^{(2,2)}$, while the other two 
are obtained from the latter two by rotating the $O{\tilde x} {\tilde z}$ plane, resulting into the simultaneous change of the sign of the multipole vectors $({\bf v}^{(2,1)}, {\bf v}^{(2,2)}) \leftrightarrow (-{\bf v}^{(2,1)}, -{\bf v}^{(2,2)})$. 
Clearly, this does not introduce any ambiguity in 
the definition of the amplitude $g_*$ and the angle $\chi$ 
(still, we allow the latter to vary within the region $0^{\circ}\leq \chi \leq 90^{\circ}$). 
This readily follows from expressions~\eqref{chimv} and~\eqref{gstarmv}, which are 
invariant under the interchange of the multipole vectors, i.e., ${\bf v}^{(2,1)} \leftrightarrow 
{\bf v}^{(2,2)}$, and under the simultaneous change of their signs.

\end{document}